\documentclass[aps, prb, 12pt, letterpaper, groupaddress]{revtex4-2}
\usepackage[utf8]{inputenc}
\usepackage[english]{babel}
\usepackage{amsmath}
\usepackage{amssymb}
\usepackage{graphicx}
\usepackage{multirow} 
\usepackage{caption}
\usepackage{subcaption} 
 \usepackage{ragged2e}

\usepackage{bm}
\usepackage[colorlinks = true,
            linkcolor = blue,
            urlcolor  = blue,
            citecolor = blue,
            anchorcolor = blue]{hyperref}

\begin{document}

\title{Characterizing nanomagnetic arrays using restricted Boltzmann machines}

\author{Rehana Begum Popy}
\email{begumpor@myumanitoba.ca}
\author{Mahdis Hamdi}
\author{Robert L. Stamps}
\affiliation{Department of Physics and Astronomy, University of Manitoba, Winnipeg, Manitoba, R3T 2N2, Canada}

\date{\today}

\begin{abstract}

Restricted Boltzmann machines are used for probabilistic learning and are capable of capturing complex dependencies in data. They are employed for diverse purposes such as dimensionality reduction, feature learning and can be used for representing and analyzing physical systems with minimal data. In this paper, we investigate a complex, strongly correlated magnetic spin system with multiple metastable states (magnetic artificial spin ice) using a restricted Boltzmann machine. Magnetic artificial spin ice is of interest because degeneracies can be specified leading to complex states that support unusual collective dynamics. We investigate two distinct geometries exhibiting different low-temperature orderings to evaluate the machine's performance and adaptability in capturing diverse magnetic behaviors. Data sets constructed with spin configurations importance-sampled from the partition function of square and pinwheel artificial spin ice Hamiltonians at different temperatures are used to extract features of distributions using a restricted Boltzmann machine. Results indicate that the restricted Boltzmann machine algorithm is sensitive to
features that define the artificial spin ice configuration space and is able to reproduce the thermodynamic quantities of the system away from criticality - a feature useful for faster sample generation. Additionally, we demonstrate how the restricted Boltzmann machine can distinguish between different artificial spin ice geometries in data even when structural defects are present. 
 
\end{abstract}

\maketitle

\section{Introduction}
\label{sub:section1}

The machine learning process usually begins by ``training" the algorithm on a large dataset, after which the trained model is tasked with performing a specific function. Many machine learning applications today utilize neural networks, which effectively map data onto a graph structure consisting of nodes and connections. For classification tasks such as image or speech recognition, networks can be trained using labelled data as in supervised learning. However, labelled data is often limited and difficult to obtain. A more effective strategy is to employ generative models that learn the entire data distribution without requiring labels (unsupervised learning). This generative approach enables the network to extract more information and even produce approximate samples from the learned distribution. This work explores the possibilities of using such an unsupervised neural network to reproduce statistical distributions of a complex magnetic system. A natural choice for this task is a restricted Boltzmann machine.

The restricted Boltzmann machine (RBM) is a generative, unsupervised, probabilistic deep-learning algorithm\cite{ackley1985learning}. It is characterized by interconnected neuron-like nodes arranged in an undirected graph structure. RBM learns a probability distribution over a set of input data and produces new data based on the learnt distribution. It is a system designed to discern and replicate patterns from a dataset. This stochastic learning algorithm is used for dimensionality reduction\cite{hinton2006reducing}, feature learning\cite{coates2011analysis}, collaborative filtering\cite{salakhutdinov2007restricted}, classification tasks\cite{hinton2006fast}\cite{larochelle2008classification} and topic modeling\cite{hinton2009replicated}. It can be a powerful tool for identifying complex features within a joint probability distribution of the system. A previous example was given by Torlai and Melko\cite{torlai2016learning} who studied the thermodynamics of the 2D Ising system using restricted Boltzmann machines. Building on this work, Yevick and Melko\cite{yevick2021accuracy} later utilized RBMs to examine the joint distribution of energy and magnetization in the 2D Ising system. Here, we use RBM as a tool to investigate a more complex system with multiple metastable states and intricate correlations between the elements called artificial spin ice\cite{skjaervo2020advances} and explore the sensitivity of RBM to different training parameters and data types. 

Artificial spin ice (ASI) systems can display an intricate and complex configuration space which arises from the interplay of the geometrical arrangement of the magnetic elements and the strength and nature of interactions between the elements. Emergent phenomena have been identified from the collective behaviors of magnetic elements. Examples include the formation of magnetic monopoles\cite{ladak2010direct}\cite{mengotti2011real}, magnetic charge ordering, vertex-based frustration\cite{gilbert2014emergent}\cite{morrison2013unhappy} and chiral dynamics\cite{gliga2017emergent}. The controllable nature of artificial spin ice holds potential for applications to data storage, microwave filtering, and energy-efficient machine learning\cite{torrejon2017neuromorphic}\cite{zhou2016large}\cite{stromberg2024design}. One key aspect of ASI is their complex relaxation dynamics and non-equilibrium behaviors due to multiple metastable states. ASI systems have been proposed as platforms for neuromorphic computing\cite{skjaervo2020advances}\cite{jensen2018computation} due to an abundance of accessible nonvolatile states, inherent nonlinearity, scalability, local interactions and rich dynamics.
 
\begin{figure}[h!]
\centerline{\includegraphics[width=0.6\textwidth]{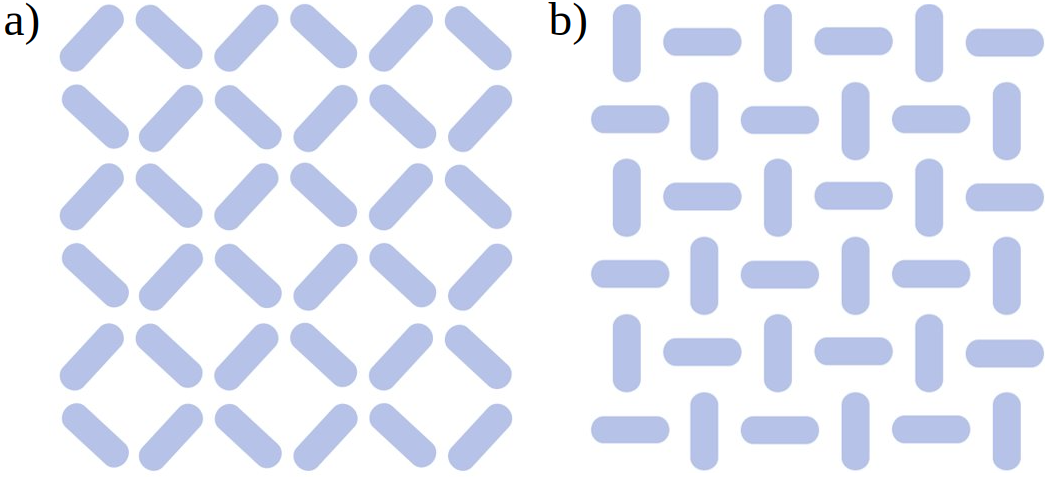}}
\caption{Two magnetic artificial spin ice geometries: a) a square and b) a pinwheel where each element is a single domain nano-magnet. Low-temperature orderings in square ASI and pinwheel ASI are antiferromagnetic and ferromagnetic respectively.}
\label{fig:fig1}
\end{figure}

Our training data is generated from Metropolis Monte Carlo sampling of two different types of artificial spin ice geometries: square ASI\cite{wang2006artificial} and pinwheel ASI\cite{gliga2017emergent}\cite{macedo2018apparent}. These geometries are illustrated in Fig. \ref{fig:fig1}. These two ASI geometries differ in how they order at low temperatures. While square ASI exhibits antiferromagnetic ordering below its ordering temperature, pinwheel ASI exhibits ferromagnetic ordering. The RBM is trained on this data set and used to produce approximate reconstructions of the original data set. A reconstruction is essentially an estimate of the probability distribution of the original input, contained in the limited-size training data set.  

The first part of this paper is dedicated to establishing the necessary models and methods required for collecting training data and training the RBM. In section \ref{sub:section3} we demonstrate RBM learning of ASI state distributions for different geometries. The sensitivity of learning to different parameters including the temperature at which the training data is generated is discussed. We explore RBM's capability of learning the statistics of the ASI system through the calculation of average system energy as a function of temperature. We show in section \ref{sub:section4} that ASI geometries above their ordering temperatures can be classified even in the presence of structural defects using trained RBMs. A summary and discussion are provided in section \ref{sub:section5}.

\section{Model and Methods}
\label{sub:section2}
For simplicity, each ASI element is assumed to be in a single magnetic domain state whose magnetization reversal approximates that of an Ising spin. The magnetic field produced by each element is approximated as a point magnetic dipole located at the element's centre. This model of interactions treats the elements as individual dipole moments and captures qualitatively the most key features of weakly coupled magnetic elements. The Hamiltonian is expressed as:

\begin{equation}
\mathcal{H}_{dip} = D \sum_{i\neq j}s_{i}s_{j} \Bigg[{\frac{{\hat{\sigma_{i}} \cdot {\hat{\sigma_{j}}}}}{r_{ij}^3}}- {\frac{3}{r_{ij}^5}}({\hat{\sigma_{i}}} \cdot {{\vec{r}}_{ij}})({\hat{\sigma_{j}}} \cdot {{\vec{r}}_{ij}})\Bigg]
\label{eqn:eqn1}
 \end{equation}
where $D=\frac{\mu_{0}\mu^2}{4\pi a^3}$ provides the energy scale. $\mu_{0}$ and $\mu$ represent the permeability constant of free space and magnetic moment strength, respectively, while $a$ is the lattice parameter taken as the nearest neighbour distance. $\mathbf{{s_{i}}}=\mu s_{i}\hat{\sigma}_{i}$ is the magnetic moment of i\textsuperscript{th} element with spin polarity $s_{i}=\pm1$, $\hat{\sigma}$'s are unit vectors parallel to the island easy axes, ${{\vec{r}}_{ij}}$ is the position vector from the i\textsuperscript{th} element to the j\textsuperscript{th} element and the subscript $i=1,2,..., N$ where $N$ is the total number of spin elements in the system. At a given temperature $T$, the configuration $\mathbf{s} = [\mathbf{s_1},\mathbf{s_2},\mathbf{s_3},...\mathbf{s_N}]^T$ drawn from the ASI sample space follows the Boltzmann distribution,

\begin{equation}
 p (\textbf{s},T) =  \frac{1}{Z_T} exp(-\frac{{\mathcal{H}_{dip}}(\mathbf{s})}{k_B T})
\end{equation}

\noindent where $Z_T = \sum\limits_{\mathbf{s}} \exp(-\frac{{\mathcal{H}_{dip}}(\mathbf{s})}{k_B T})$ is the partition function and $k_B$ is the Boltzmann constant.

The RBM is trained on data generated by $300,000$ successive configurations of a $6\times6$ finite ASI system using Markov chain Metropolis Monte Carlo sampling at a temperature $T$ in units of $D/k_B$. The results discussed throughout this paper are obtained by training either a portion of this data or the full set. Before collecting the data, the system is equilibrated by running $15,000$ Monte Carlo steps on the system. In addition to generating sample configurations, MC simulations also compute estimates of thermodynamic properties, such as energy and magnetization, by directly analyzing the sampled probability distribution of the system under study. These estimators are also obtained from the RBM reconstructed approximate distributions, which are compared with those obtained from the original training data set.

We use an RBM consisting of two layers of units. One layer is visible to the outside world and represents observable data. The other layer is hidden and connects only to the visible nodes. Every connection has an associated weight $w_{ij}$ and each visible and hidden unit has a bias ascribed to it. Units within the same layer are not connected. A diagram of the network geometry is given in Appendix \ref{ap:ap1}.

The energy function of the RBM has an Ising-like form\cite{brush1967history} and is defined as:
\begin{equation}
 E(v,h) = - \sum_{i} a_i v_i - \sum_{j} b_j h_j - \sum_{i} \sum_{j} v_i W_{i,j} h_j
\label{eqn:eqn2}
\end{equation}
where $W_{i,j}$ is the weight of the interaction between visible unit $v_i$ and hidden unit $h_j$, $a_i$ and $b_j$ are biases for $v_i$ and $h_j$ respectively, $i=1,2,..., n_v$ and  $j=1,2,..., n_h$ where $n_v$ and $n_h$ are the numbers of visible and hidden nodes respectively.
The RBM assumes probabilities of the following form:

\begin{equation}
 p_{\lambda}(v,h) =  \frac{1}{Z} exp(-E(v,h))
\label{eqn:eqn3}
\end{equation}
where $Z$ is the normalization term and $\lambda = \{\mathbf{W}, \mathbf{a},\mathbf{b}\}$ are the training parameters. The goal of RBM training is to build a probabilistic model $p_{\lambda}(v,h)$ which mimics the target distribution $p (\textbf{s}, T)$ by tuning the training parameter set $\lambda$. The optimum training parameters will minimize the distance between these two distributions. RBM training aims to maximize the expected log probability assigned to a given training set by minimizing the energy function. In this paper, we use a combination of Gibbs sampling\cite{geman1984stochastic} and contrastive
divergence($\text{CD}_k$)\cite{hinton2002training} to maximize the expected log probability of the visible vector $v$. Details of the RBM training process are included in Appendix \ref{sub:subsection5}.

We use an RBM consisting of 36 visible units chosen to match the number of spins in a $6\times 6$ lattice. To mitigate the risk of overfitting commonly linked to limited training data, we constructed a sufficiently large dataset comprising 300,000 samples. Also, to allow matrix-matrix GPU multiplication to be used for substantial speed up in the learning process, the training set is divided into small batches of $50$. We initialize the biases in Eq. \ref{eqn:eqn2} to zero and select initial weights from a uniform distribution around zero and width $\propto \sqrt{1/(n_v +n_h)}$. Through trial and error, we determine two critical parameters: the number of hidden nodes and the learning rate. Following Hinton's guidelines\cite{hinton2012practical} and additional experimentation, our tests indicate that 36 hidden nodes provide optimal learning performance based on training errors. The learning rate determines the size of the steps taken towards minimizing the error during optimization\cite{hinton2012practical}.  A learning rate that is too high can cause the model to converge to a sub-optimal solution, while a rate that is too low may result in a lengthy training process that may not converge to the best solution. A learning rate of $0.0001$ is determined as optimal and the training process is executed for a total of 15,000 epochs. We find that $\text{CD}_5$ is sufficient to reconstruct the visible states with low error. Training errors are computed from the sum of squared errors (obtained from the element-wise difference between the input vector and RBM reconstruction vector) averaged over the batches. 

Based on the ASI geometry depicted in Fig. \ref{fig:fig1}, it is evident that the spin elements in the ASI system are vectors capable of assuming any of the four distinct orientations. To train the RBM on the spin configurations of the ASI system, these vectors are transformed into binary units (0,1) according to a predetermined sign convention. The training data is arranged randomly using a permuted index vector to reduce the correlation between subsequent Metropolis realizations. The trained RBM with learnt weights and biases can be used to reconstruct the training state distribution. The reconstructions from random input can be used as independently sampled data that replicate macroscopic quantities of interest such as the energy and magnetization of the ASI system.

\section{Data distribution learning with RBM} 
\label{sub:section3}

\subsection{Square ASI }
\label{sub:subsection1}
We now examine the distribution of states using RBM for a square ASI geometry. A configuration state is mapped onto system energy and in-plane magnetization components. The joint distribution constructed from all configurations is consistent with each mapped energy-magnetization pair. Only the in-plane magnetization along the x-axis is shown in the figures as the magnetization along the y-axis has a similar response and does not add any additional information. 

The RBM is sensitive to the size of the training data, training parameters and the temperature at which the training data is generated. Since the goal of the RBM training process is to approximate the original training distribution, a natural choice for comparing the model's learned distribution to the actual data distribution is the Kullback-Leibler (KL) divergence. The Kullback-Leibler divergence is an information-theoretic measure that quantifies the difference between two probability distributions\cite{kullback1951information}. The KL divergence is always non-negative and a value of 0 indicates that the training data and their reconstruction are identical.  

While the geometry of the ASI restricts the system from exhibiting only discrete magnetization values, the system energy has a continuous form. Accordingly, a mixed discrete and continuous definition is used for the KL divergence:

\begin{equation}
   \text{KL}(P||Q) = \sum_{x} \int_{y} P(x,y) ln \frac{P(x,y)}{Q(x,y)} dy
\label{eqn:eqn4}
\end{equation}

\noindent where the summation is over all possible discrete values of the x-component of magnetization and the integration is over the entire domain of the continuous variable y representing the energy of the system.

\begin{figure}[!tbp]
  \centering
   \subfloat[][\centering]{\includegraphics[width=0.498\textwidth]{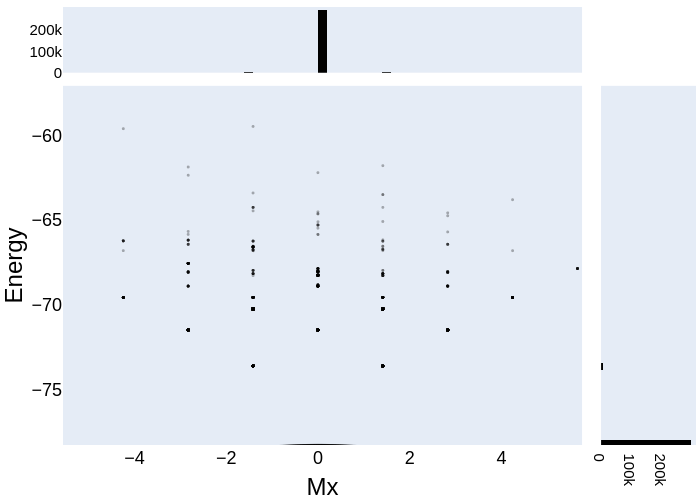}}
  \hfill
  \subfloat[][\centering]{\includegraphics[width=0.498\textwidth]{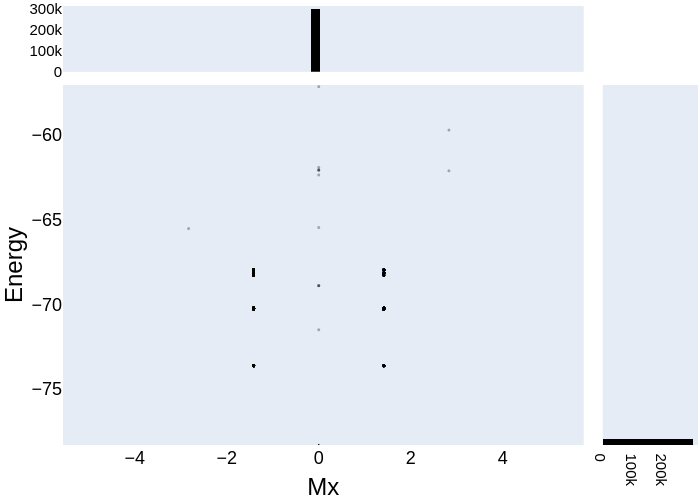}}
  \caption{2D histogram contour plots showing the square artificial spin ice states in energy-magnetization space. (a) Distribution from $300,000$ Monte Carlo realizations at $T=0.9$ $D/k_B$. (b) Distribution from $300,000$ samples generated by an RBM trained on Monte Carlo data at the same temperature. Insets show frequency histograms of energy and magnetization values for each dataset. The similarity between (a) and (b) demonstrates the RBM's ability to accurately reconstruct the low-temperature state distribution of the square ASI system.}
  \label{fig:fig2}
\end{figure}

Fig. \ref{fig:fig2} shows the efficacy of RBM at a low-temperature regime ($T = 0.9$ $D/k_B$) by comparing the joint energy-magnetization distribution plot derived from the Monte Carlo simulations (Fig. \ref{fig:fig2}(a)) and RBM reconstruction (Fig. \ref{fig:fig2}(b)). The results are obtained after training the RBM on $300,000$ Monte Carlo simulated square ASI configurations and subsequently testing it on $300,000$ randomly generated configurations using the learned weights and biases. Included are 2D histogram contour plots showing the distribution of data points across the energy and x-axis magnetization with contours that represent equal probability superimposed on a scatter plot displaying individual data points. The histograms show distributions summed over energy (side) and summed over magnetization (top).

Fig. \ref{fig:fig2}(a) reveals a striking concentration of states in a very localized region of the configuration space, characterized by a magnetization value of zero and a very low energy value. This dense clustering appears as prominent, intense peaks in both axes, indicating that the vast majority of ASI states occupy this narrow range of magnetization and energy values. This observation aligns with theoretical expectations: at low temperatures, the most probable state is the ground state of the system, and for square ASI, the ground state corresponds to the one with the lowest energy and zero net magnetization, reflecting its antiferromagnetic nature. The RBM reconstruction (Fig. \ref{fig:fig2}(b)) successfully captures these essential features, exhibiting single, distinct peaks along both the energy and magnetization axes that closely mirror the original data. This accurate reproduction highlights the RBM's capability to effectively model the distribution of low-temperature states in the square ASI system. Furthermore, this performance is quantitatively validated by a KL divergence of $0.001$, indicating an excellent agreement between the Monte Carlo-generated data and the RBM reconstruction.

\begin{figure}[!tbp]
  \centering
  \subfloat[][\centering]{\includegraphics[width=0.498\textwidth]{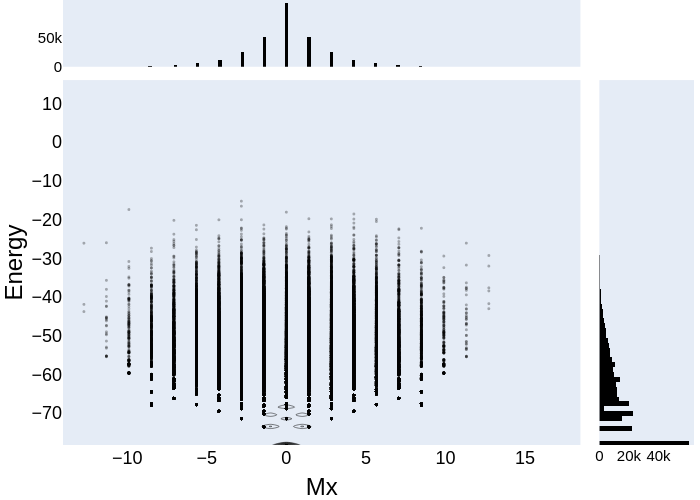}}
  \hfill
  \subfloat[][\centering]{\includegraphics[width=0.498\textwidth]{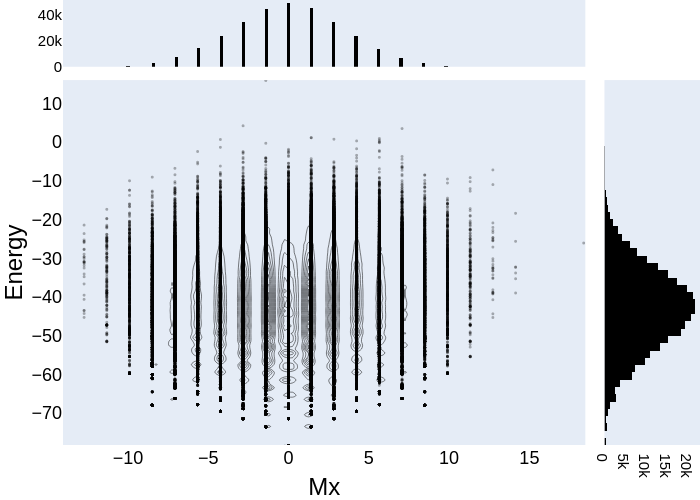}}
  \caption{2D histogram contour plots showing the square artificial spin ice states in energy-magnetization space. (a) Distribution from $300,000$ Monte Carlo realizations at $T=2.0$ $D/k_B$. (b) Distribution from $300,000$ samples generated by an RBM trained on Monte Carlo data at the same temperature. Insets show frequency histograms of energy and magnetization values for each dataset. The difference between (a) and (b) demonstrates the RBM's failure to accurately capture the details of the state distribution of the square ASI system in the critical region.}
  \label{fig:fig3}
\end{figure}

In contrast to its success at low temperatures, the RBM's performance in reconstructing the state distribution of the square ASI system deteriorates significantly as we approach the ordering temperature, $T_c$. The ordering temperature of square ASI for our geometry is $T_c = 2.54$ $D/k_B$, consistent with values reported in\cite{silva2012thermodynamics}. Fig. \ref{fig:fig3} illustrates this limitation by comparing the energy and magnetization distributions from $300,000$ Monte Carlo samples generated at $T = 2.0$ $D/k_B$ with those produced by the RBM trained on this data. In the MC sampled data (Fig. \ref{fig:fig3}(a)), the energy distribution exhibits a distinct skew and is concentrated at very low values, reflecting the system's tendency to occupy lower energy states. In contrast, the RBM reconstruction (Fig. \ref{fig:fig3}(b)) produces a Gaussian distribution with a notably larger width and a peak shifted towards higher energy values. This suggests that the RBM fails to capture the asymmetry and fine structure of the energy landscape near the ordering temperature. The magnetization distribution shows similar disparities. The MC-generated data displays a sharper peak, indicating a more defined preferential magnetization state. The RBM reconstruction, however, yields a broader, more diffuse distribution, underestimating the system's tendency to maintain specific magnetization values. These qualitative observations are quantitatively supported by a KL divergence of $0.093$ between the MC and RBM distributions. This relatively high KL divergence value with respect to the low-temperature value signifies a substantial discrepancy, highlighting the RBM's struggle to accurately model the system's behavior in this temperature regime.

\begin{figure}[!tbp]
  \centering
  \subfloat[][\centering]{\includegraphics[width=0.498\textwidth]{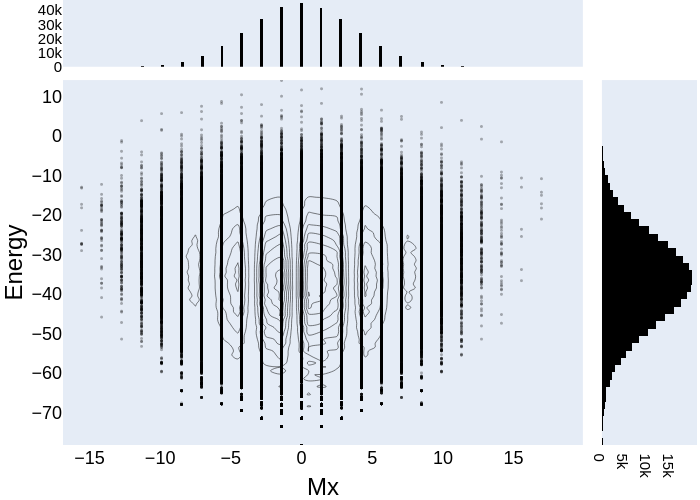}}
  \hfill
  \subfloat[][\centering]{\includegraphics[width=0.498\textwidth]{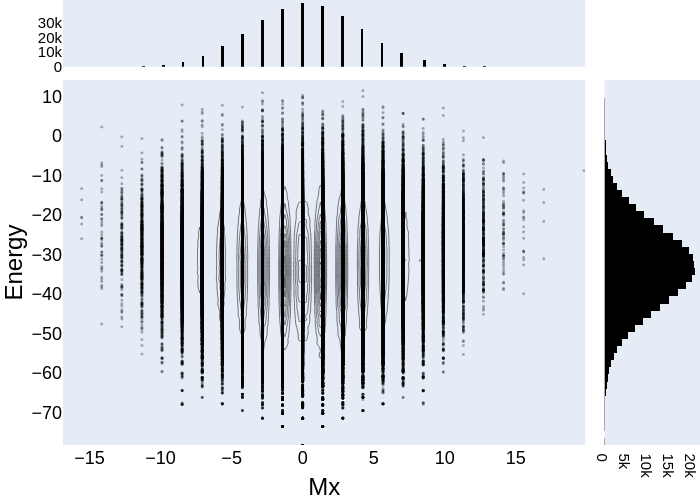}}
  \caption{2D histogram contour plots showing the square artificial spin ice states in energy-magnetization space. (a) Distribution from $300,000$ Monte Carlo realizations at $T=3.5$ $D/k_B$. (b) Distribution from $300,000$ samples generated by an RBM trained on Monte Carlo data at the same temperature. Insets show frequency histograms of energy and magnetization values for each dataset. The similarity between (a) and (b) demonstrates the RBM's ability to accurately reconstruct the high-temperature state distribution of the square ASI system.}
  \label{fig:fig4}
\end{figure}

Fig. \ref{fig:fig4} illustrates the RBM's performance at a high temperature of $T = 3.5$ $D/k_B$, which is significantly higher than the ordering temperature. In contrast to the discrepancies observed near $T_c$, here we see a good agreement between the Monte Carlo data and the RBM reconstruction. Both the energy and magnetization distributions show nearly identical shapes, peaks and ranges. The energy distribution exhibits a symmetric, Gaussian-like profile characteristic of high-temperature behavior, which the RBM reproduces with high fidelity. Similarly, the magnetization distribution shows a broader, more uniform spread, reflecting the increased randomness at high temperatures, which is accurately captured by the RBM. This is supported by a low KL divergence of $0.005$, comparable to the performance observed at low temperatures.

\begin{figure}[h!]
\centerline{\includegraphics[width=0.6\textwidth]{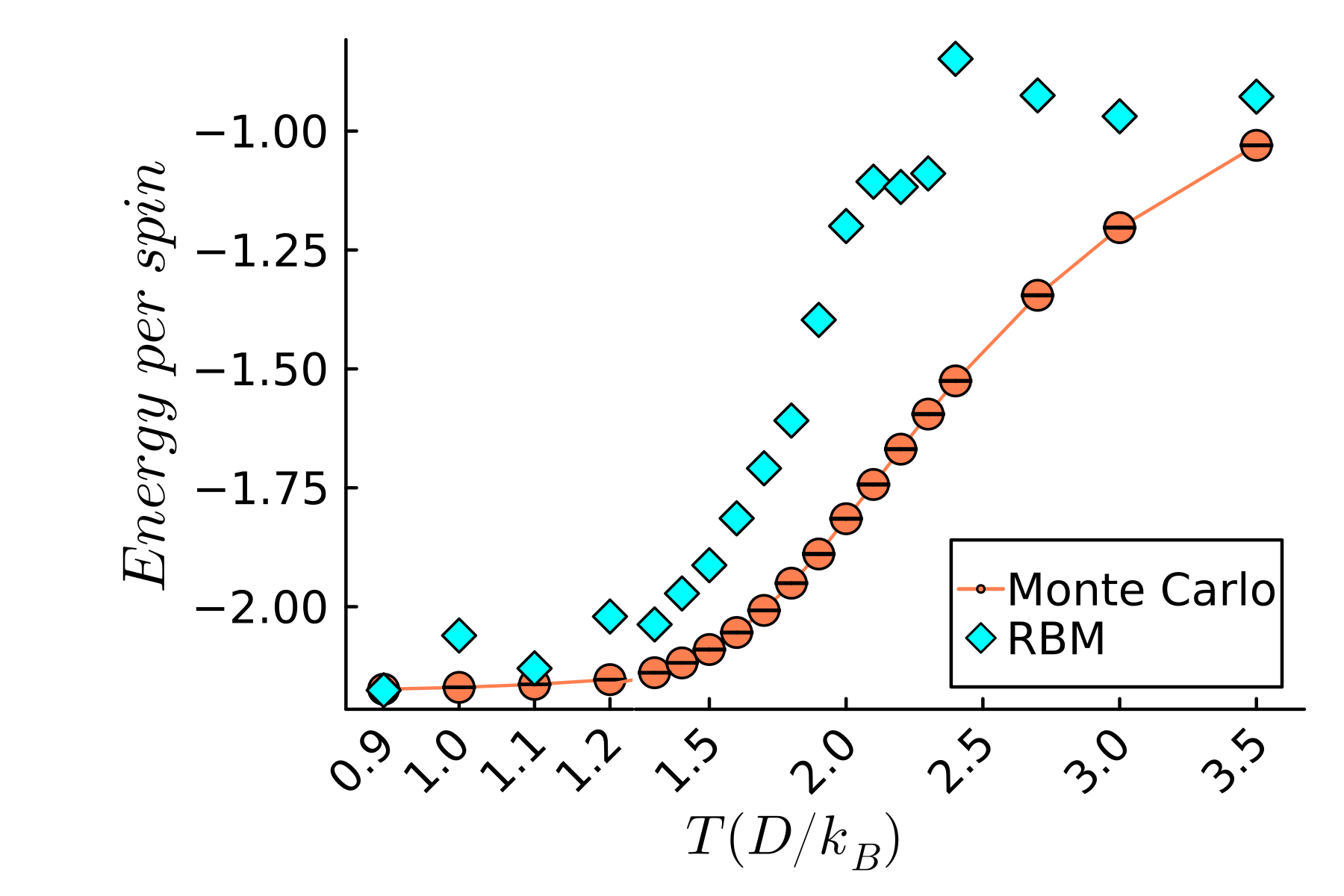}}
\caption{The average energy per spin of a $6\times6$ square ASI as a function of temperature obtained from Monte Carlo generated data and RBM reconstructed data. With the exceptions near criticality, RBM reconstruction seems to follow a similar trend to that of the MC samples. The error bars in Monte Carlo generated data are obtained over five independent trials. For RBM reconstructions, the error bars are negligible and not shown here.}
\label{fig:fig5}
\end{figure}

We now examine how well the RBM captures the thermodynamic properties of the square ASI system across different temperature regimes. Fig. \ref{fig:fig5} illustrates the average system energy per spin as a function of temperature, comparing results from Monte Carlo simulations with those obtained from RBM reconstructions. The Monte Carlo data, serving as our reference, is generated from simulations of a $6\times6$ square ASI lattice. For each temperature point, we train a separate RBM on $300,000$ Monte Carlo generated samples, ensuring that each RBM accurately represents the system's state at that specific temperature.  At low temperatures, the RBM reconstructions closely match the Monte Carlo data, accurately capturing the highly ordered system state. Similarly, at high temperatures, the RBM demonstrates an acceptable agreement, effectively modelling the random configurations. However, near $T_c=2.54$ $D/k_B$, in the critical region, significant deviations emerge, reflecting the RBM's struggle in capturing the complex correlations and fluctuations characteristic of the phase transition. Despite these discrepancies in the critical region, the RBM successfully reproduces the overall energy-temperature relationship, suggesting its ability to learn and represent the fundamental thermodynamic properties of the square artificial spin ice system across most of the temperature range.

\subsection{Pinwheel ASI}
\label{sub:subsection2} 
Pinwheel artificial spin ice exhibits ferromagnetic ordering at low temperatures, characterized by the formation of various types of magnetic domains and domain walls\cite{macedo2018apparent}. A recent study has indicated that the ordering temperature for pinwheel ASI is approximately $T_c \approx 0.7$ $D/k_B$\cite{popy2022investigation} for a $10\times10$ pinwheel ASI lattice.


As observed with square ASI, the RBM is able to capture the state distribution of pinwheel ASI at low temperatures. Specifically, at $T=0.06$ $D/k_B$, which is well below the ordering temperature, the RBM reconstructs the detailed features of the pinwheel ASI state distribution (results not shown here). The high fidelity of this reconstruction is confirmed by a KL divergence of $0.002$ between the Monte Carlo-generated data and the RBM reconstruction at this temperature.

\begin{figure}[!tbp]
  \centering
  \subfloat[][\centering]{\includegraphics[width=0.5\textwidth]{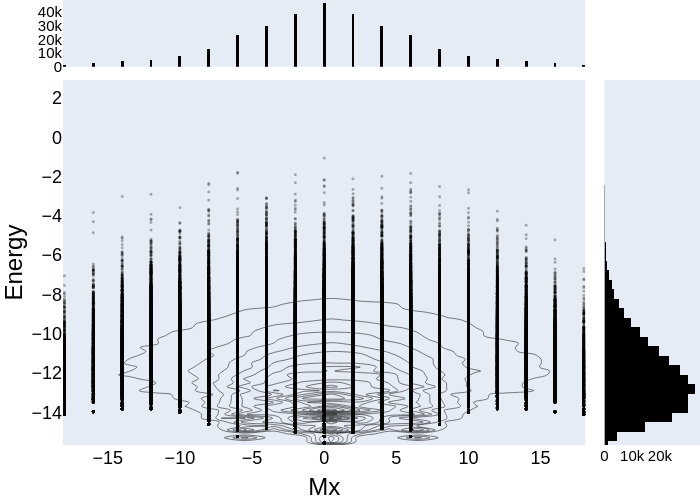}}
  \hfill
  \subfloat[][\centering]{\includegraphics[width=0.5\textwidth]{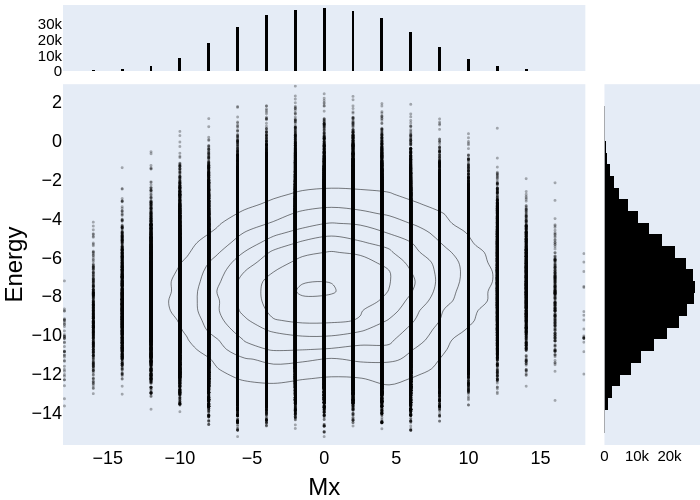}}
  \caption{2D histogram contour plots showing the pinwheel artificial spin ice states in energy-magnetization space. (a) Distribution from $300,000$ Monte Carlo realizations at $T=0.5$ $D/k_B$. (b) Distribution from $300,000$ samples generated by an RBM trained on Monte Carlo data at the same temperature. Insets show frequency histograms of energy and magnetization values for each dataset. The difference between (a) and (b) demonstrates the RBM's failure to accurately capture the details of the state distribution of the pinwheel ASI system in the critical region.}
  \label{fig:fig6}
\end{figure}

\begin{figure}[!tbp]
  \centering
  \subfloat[][\centering]{\includegraphics[width=0.5\textwidth]{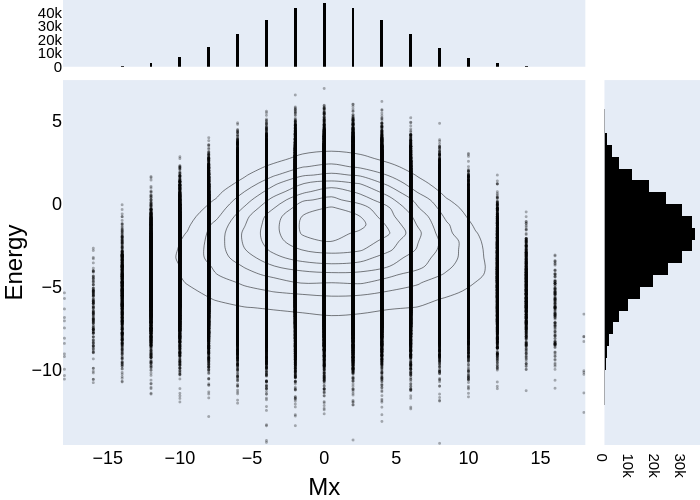}}
  \hfill
  \subfloat[][\centering]{\includegraphics[width=0.5\textwidth]{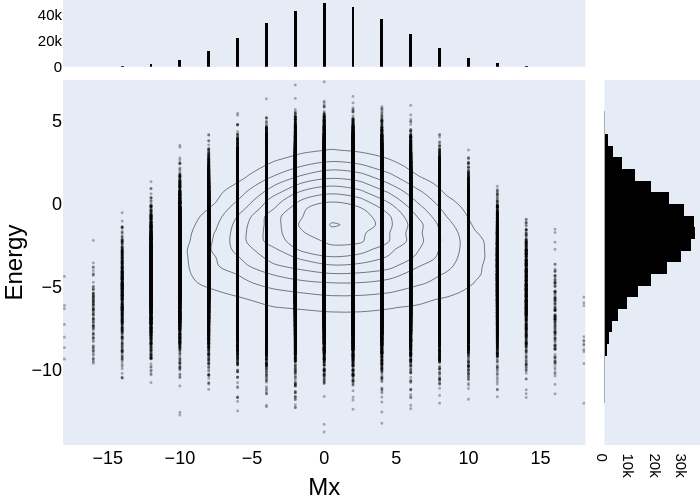}}
  \caption{2D histogram contour plots showing the pinwheel artificial spin ice states in energy-magnetization space. (a) Distribution from $300,000$ Monte Carlo realizations at $T=2.5$ $D/k_B$. (b) Distribution from $300,000$ samples generated by an RBM trained on Monte Carlo data at the same temperature. Insets show frequency histograms of energy and magnetization values for each dataset. The similarity between (a) and (b) demonstrates the RBM's ability to accurately reconstruct the high-temperature state distribution of the pinwheel ASI system.}
  \label{fig:fig7}
\end{figure}

However, as we approach the critical region, the RBM's performance deteriorates significantly, mirroring the behavior observed in square ASI. Fig. \ref{fig:fig6} illustrates this limitation by comparing the energy and magnetization distributions from $300,000$ Monte Carlo samples generated at $T=0.5$ $D/k_B$ with those produced by the RBM trained on this data. The energy distribution obtained from the MC sampled training data is skewed towards lower energy values, indicating a preference for low-energy states that reflect the system's tendency to occupy energetically favourable configurations. In contrast, the RBM produces a Gaussian distribution for energy, with its peak shifted towards higher energy values, suggesting that it fails to capture the inherent asymmetry of the energy landscape. Similarly, the magnetization values from RBM reconstruction yield a distribution with identical mean but larger variance, which implies an overestimation of variability in magnetization states. This discrepancy in the overall state distribution of the system is evident from a relatively larger KL divergence of $0.215$. Interestingly, as we move to temperatures well above the critical point, the RBM's performance in reconstructing the state distribution of pinwheel ASI improves, similar to the case of square ASI (see Fig. \ref{fig:fig7}). Energy and magnetization histogram plots have Gaussian forms with similar means and moments. 
In this case, the KL divergence between the MC training data and RBM reconstruction at $T=2.5$ $D/k_B$ (high-temperature regime) is calculated to be $0.003$ which is comparable to the value of $0.002$ calculated at $T=0.06$ $D/k_B$ (low-temperature regime).

\begin{figure}[h!]
\centerline{\includegraphics[width=0.6\textwidth]{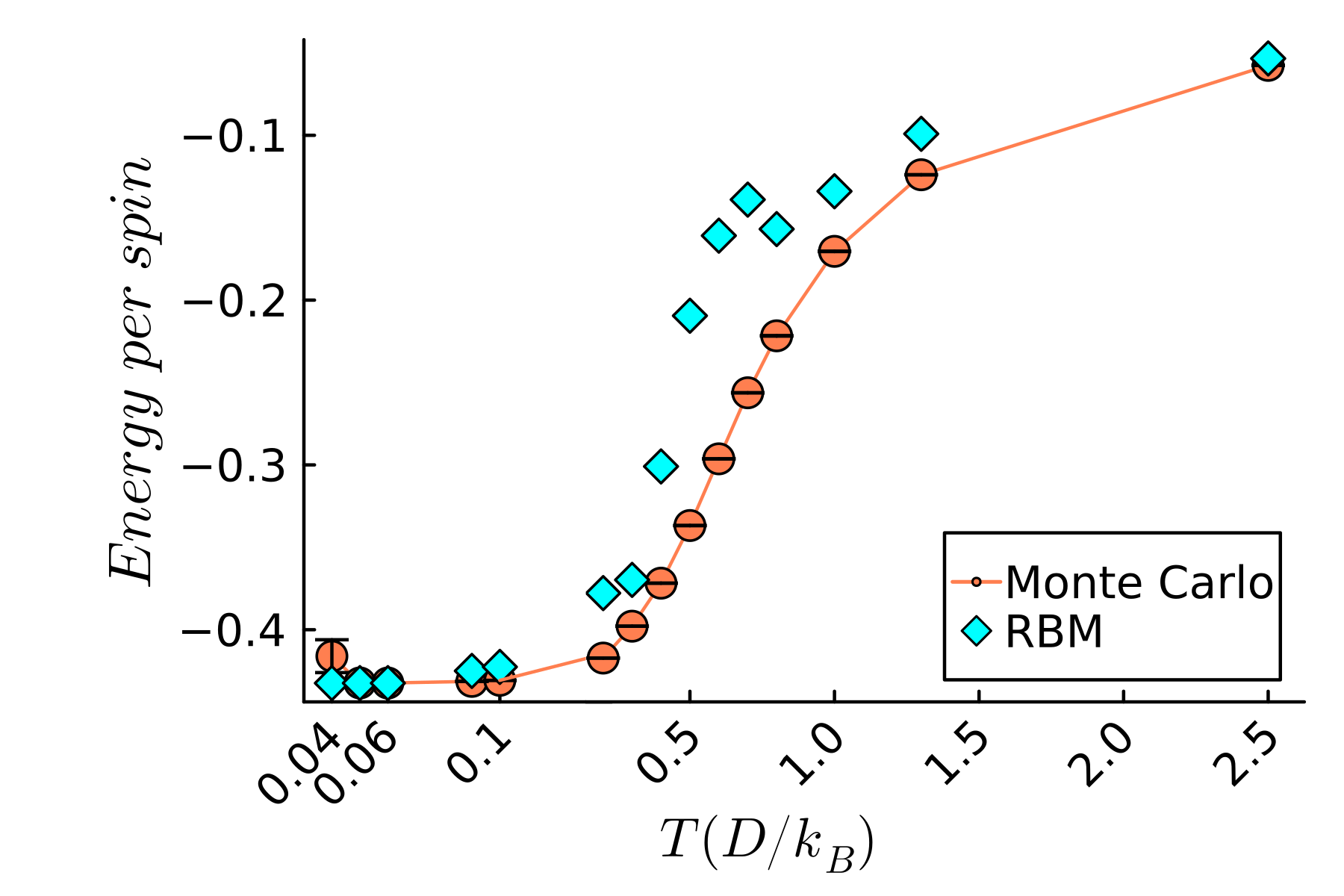}}
\caption{The average energy per spin of a $6\times6$ pinwheel ASI as a function of temperature obtained from Monte Carlo generated data and RBM reconstructed data. With the exceptions near criticality, RBM reconstruction seems to follow a similar trend to that of the MC samples. The error bars in Monte Carlo generated data are obtained over five independent trials. For RBM reconstructions, the error bars are negligible and not shown here.}
\label{fig:fig8}
\end{figure}

The graph in Fig. \ref{fig:fig8} depicts the average system energy per spin as a function of temperature for pinwheel ASI, comparing results from Monte Carlo simulations with those obtained from RBM reconstructions. The RBM's performance for pinwheel ASI is comparable with that of square ASI in the low and high-temperature regimes, showing good agreement with Monte Carlo data. However, there are pronounced differences in the critical region. The deviations between RBM and Monte Carlo results appear less pronounced in pinwheel ASI, compared to the square ASI. We note that the pinwheel ASI has different ordering mechanisms and emergent domain structures\cite{macedo2018apparent}. Near the critical temperature, square ASI exhibits energy and magnetization distributions tightly concentrated in a narrow region of configuration space. In contrast, pinwheel ASI develops complex domain and domain wall configurations, resulting in a more expansive and varied distribution of energy and magnetization values. The broader, more distributed configuration space of pinwheel ASI may facilitate RBM's ability to extract underlying statistical properties.

These observations suggest that the RBM performs best (both for the square and pinwheel ASI geometries) in regimes far from criticality, both at low and high temperatures. In these regimes, the system's behavior is dominated by either highly ordered states (low T) or largely random configurations (high T), which the RBM can effectively model. The challenge for the RBM appears to be specifically in the critical regime near $T_c$, where complex, long-range correlations and fluctuations dominate the system's behavior. Torlai and Melko\cite{torlai2016learning} investigated the performance of RBMs in modelling the 2D Ising model near criticality. Their findings revealed that at the critical temperature, a significantly larger number of hidden units was required to accurately capture the thermodynamics of the system. For ASI systems, the effect of increasing the number of hidden nodes at criticality needs a thorough investigation which is not the goal of this paper.


\section{Data classification using RBM} 
\label{sub:section4}
In addition to modelling the probability distribution of data and generating new samples from that distribution, restricted Boltzmann machines have demonstrated utility in classification tasks. They have been predominantly utilized either to initialize the weight of neural network classifiers or as feature extractors for other classification techniques\cite{hinton2007recognize}\cite{gehler2006rate}. Subsequently, a self-contained discriminative RBM framework for creating competitive classifiers was proposed\cite{hinton2006fast}\cite{larochelle2008classification}.

We now investigate the efficacy of RBMs, in conjunction with a classifier, in categorizing two distinct artificial spin ice geometries: square and pinwheel. This is accomplished without structural modifications to the RBM. However, rather than training the RBM on a single ASI geometry, we utilize a heterogeneous dataset comprising both square and pinwheel ASI configurations. The training dataset encompasses states distributed across two separate configuration spaces: $30,000$ square ASI data and $30,000$ pinwheel data generated by the Metropolis Monte Carlo sampling algorithm at a temperature $T = 3.5$ $D/k_B$ for $6\times6$ lattices. To mitigate potential bias, the square and pinwheel data are randomly ordered.

To evaluate the trained RBM's performance, we employ four distinct classes of test datasets, departing from the previously used random data: 1) perfect square ASI data, 2) perfect pinwheel ASI data, 3) imperfect square ASI data and 4) imperfect pinwheel ASI data. The imperfect ASI configurations are utilized to assess the RBM's ability to discriminate between square and pinwheel geometries under non-ideal conditions.

In the simulation, the magnetic moment (and consequently, the strength of each ASI element) is set to unity. To construct an imperfect artificial spin ice, some of the spin elements are designed to interact weakly with others by randomly choosing either $4$ or $8$ spins out of the $36$ spins and setting their magnetic moment values to $0.5$. The classification task is accomplished through a two-step process: a) calculating KL divergence values between the RBM reconstructed state distributions and the original square and pinwheel ASI state distributions, b) identifying the type of RBM reconstruction based on these KL divergence values, determining whether the reconstructed distribution more closely resembles that of a square or pinwheel ASI.

\subsection{Testing RBM on perfect ASI data} 
\label{sub:subsection3}

In our initial trial, we evaluate the trained RBM on $60,000$ successive Metropolis Monte Carlo realizations of a perfect $6\times 6$ square artificial spin ice. As anticipated, the RBM reconstruction exhibits a distribution similar to that of the perfect square ASI data. This observation is quantitatively verified through KL divergence measurements. The comparison between the reconstructed data distribution and the original square ASI data distribution yielded a KL divergence value of $0.004$. In contrast, when comparing the reconstructed data distribution with the original pinwheel ASI data distribution, the KL divergence measurement resulted in a value of $0.123$. These quantitative results provide strong evidence that the reconstructed data possesses a state distribution that closely aligns with that of the square ASI.

Similarly, the trained RBM is tested on $60,000$ successive Metropolis Monte Carlo realizations of a perfect $6\times 6$ pinwheel artificial spin ice and the reconstruction has a closer match with the original pinwheel ASI data distribution. For these two distributions, the KL divergence value is $0.011$, which is much smaller than the KL divergence value of $0.642$ calculated for the comparison between the distributions for the square ASI data and RBM reconstructed data. Table \ref{tab:table10} summarizes the KL divergence values for different comparisons.

\begin{table}[h]
\caption{KL divergence values for testing RBM on perfect data}
\centering
\begin{tabular}{|c|c|c|c|}  
\hline
\textbf{RBM trained} & \textbf{RBM tested} & \textbf{Reconstruction} & \textbf{KLD} \\
\textbf{on perfect:} & \textbf{on perfect:} & \textbf{compared with perfect:} & \\
\hline
\multirow{4}{*}{\begin{tabular}[c]{@{}c@{}}square\\ + \\ pinwheel\end{tabular}} 
    & \multirow{2}{*}{square} & square & 0.004 \\
    & & pinwheel & 0.123 \\ 
\cline{2-4}
    & \multirow{2}{*}{pinwheel} & square & 0.642 \\
    & & pinwheel & 0.011 \\ 
\hline
\end{tabular}
\label{tab:table10}
\end{table}

\subsection{Testing RBM on imperfect ASI data} 
\label{sub:subsection4}

The artificial spin systems used in experiments are not perfect and include irregularities in the arrangement of nanomagnets, variations in the size or shape of individual nanomagnets, structural imperfections and impurities. To simulate these real-world conditions, we introduce controlled imperfections in the form of structural defects into our model. Specifically, we randomly select  $4$ spin elements out of $36$ to have half the strength of other elements. This modification alters the inter-element interactions and consequently modifies the ASI state distribution.
 
We employ Metropolis Monte Carlo sampling to generate $60,000$ successive realizations each for an imperfect $6\times6$ square ASI and an imperfect $6\times6$ pinwheel ASI. The restricted Boltzmann machine, previously trained on $30,000$ Monte Carlo simulated perfect square ASI configurations and $30,000$ perfect pinwheel configurations, is then evaluated using these imperfect datasets. 

When evaluated on imperfect square ASI data, the RBM reconstruction exhibits a state distribution remarkably similar to that of the original square ASI data distribution, yielding a relatively small KL divergence value of $0.024$. This result is particularly significant as it demonstrates the RBM's ability to successfully capture underlying features and relate them to the appropriate training distribution, despite the test data occupying a slightly different configuration space due to introduced imperfections. Similarly, when the trained RBM is tested on $60,000$ imperfect pinwheel ASI samples, it, in conjunction with the classifier, successfully identifies the test samples as having a pinwheel ASI-like data distribution. In this case, the KL divergence value is $0.014$, which is substantially smaller than the KL divergence value of $0.662$ obtained when comparing the reconstructed distribution with the original square ASI data distribution. Table \ref{tab:table20} provides a comprehensive summary of the KL divergence measurements for imperfect data with $4$ weak spins.

\begin{table}[h]
\caption{KL divergence values for testing RBM on imperfect data with $4$ and $8$ weak spins}
\centering
\begin{tabular}{|c|c|c|c|c|}  
\hline
\textbf{RBM trained} & \textbf{RBM tested} & \textbf{Reconstruction} & \textbf{KLD} & \textbf{KLD}\\
\textbf{on perfect:} & \textbf{on imperfect:} & \textbf{compared with perfect:} & \textbf{(4 defects)} & \textbf{(8 defects)} \\
\hline
\multirow{4}{*}{\begin{tabular}[c]{@{}c@{}}square\\ + \\ pinwheel\end{tabular}} 
    & \multirow{2}{*}{square} & square & 0.024 & 0.037\\
    & & pinwheel & 0.086 & 0.074\\ 
\cline{2-5}
    & \multirow{2}{*}{pinwheel} & square & 0.662 & 0.678\\
    & & pinwheel & 0.014 & 0.013\\ 
\hline
\end{tabular}
\label{tab:table20}
\end{table}

The experiment is extended to introduce more significant imperfections in the ASI systems by reducing the strength of 8 out of 36 spin elements to half that of the others. This modification further alters the inter-element interactions and the ASI state distribution. The results of this experiment are summarized in Table \ref{tab:table20}. The KL divergence values indicate that the RBM successfully identifies both imperfect square and pinwheel ASI configurations, despite the increased defects. For imperfect square ASI data with 8 weak spins, the RBM correctly identifies the square ASI-like distribution, evidenced by a relatively low KL divergence value of $0.037$. Likewise, when tested on imperfect pinwheel ASI configurations, the RBM accurately recognized the pinwheel-like distribution, yielding a comparably low KL divergence value of $0.013$.

These results suggest the RBM's robustness in classifying ASI geometries even under substantial structural modifications, highlighting its potential for extracting information about defects in ASI systems with imperfections which is not demonstrated in this work.

\section{Conclusion}
\label{sub:section5}

In this work, RBMs have been shown to successfully learn the temperature-dependent state distributions from two different artificial spin ice geometries. They learn to encode abstract features of the input, which can represent higher-order correlations and dependencies in the data. This allows the trained RBMs to generate new samples that accurately reflect the statistical properties of the system, enabling further analysis and exploration of the ASI behavior. In practical experiments, it is crucial to sample a large number of states to study the statistical properties of the system accurately, which can sometimes be time-consuming. RBM can be useful in this regard as once trained, it can faithfully reproduce new samples from the original configuration space in a significantly short time.  

However, in the vicinity of the critical region, RBM demonstrates significant limitations in accurately describing the probability distributions for both square and pinwheel ASI systems, at least within the constraints of our chosen training parameters. A key discrepancy emerges in the RBM's tendency to generate Gaussian distributions of states, whereas the original Monte Carlo data exhibits asymmetric distributions. This suggests that near criticality, RBM struggles to fully capture and reproduce the complex, temperature-dependent spin-spin correlations that are inherent in the training data. According to Torlai and Melko\cite{torlai2016learning}, it might still be possible to obtain an accurate approximation of the physical distribution of the ASI system by adjusting the number of hidden nodes. Furthermore, other training parameters need investigation, as RBM learning is sensitive to various factors including the number of hidden nodes, learning rate, number of training epochs, and Gibbs sampling steps. These avenues for optimization suggest that there may be potential for improving the RBM's performance in modelling critical phenomena in ASI systems. 

Another potential application of RBMs explored in this paper lies in correctly identifying features in the ASI distributions and employing them to classify unknown ASI samples. We demonstrate that RBMs can distinguish between different ASI geometries in mixed data, even with defect concentrations up to 22.22\%. Notably, when presented with imperfect test data, RBMs can still correctly identify characteristic features, producing reconstructions that closely mimic the perfect data distribution for the same class. This capability has significant practical implications, as experimental realizations of ASI inevitably contain defects or disorders such as irregularities, dislocations, edge boundaries, and impurities. Furthermore, by comparing RBM reconstructions trained on perfect ASI data with those trained on imperfect (defected) ASI data, it may be possible to infer the distribution of defects present in the imperfect system. This approach could provide additional insights into the nature and extent of defects in experimental ASI samples, potentially aiding in the refinement of fabrication techniques and the understanding of defect impacts on ASI properties.

\begin{acknowledgements}
This work was supported by The Natural Sciences and Engineering Research Council of Canada (NSERC) Discovery, John R. Leaders Fund - Canada Foundation for Innovation (CFI-JELF), Research Manitoba and the University of Manitoba, Canada.
\end{acknowledgements}

\bibliographystyle{unsrt}
\bibliography{main}

\section{Appendix }
\label{ap:ap1}

\subsection{Training RBM}
\label{sub:subsection5}

\begin{figure}[h!]
\centering
\includegraphics[width=0.4\linewidth]{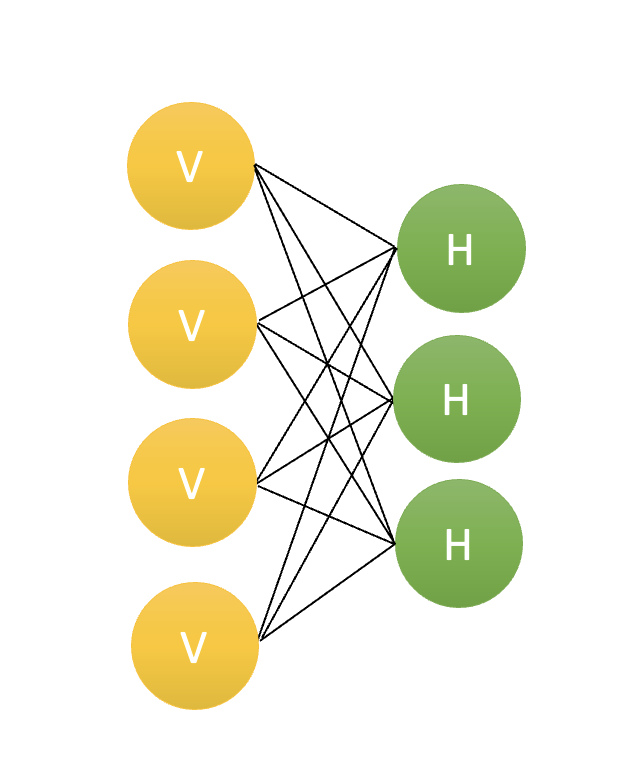}
\caption{Schematic representation of a restricted Boltzmann machine. Yellow circles are visible units in the first layer. Green circles are probabilistic hidden units in the second layer.}
\label{fig:fig9}
\end{figure}

The neurons in the hidden layer receive input data from neurons in the visible layer. The weights are multiplied by the inputs, and the bias is added. The resulting product goes through the activation function. The output of the activation function, which is subsequently applied to the result, determines whether or not the hidden state is activated. Multiple inputs are combined and contribute to the activation of a single hidden node. This process is repeated for all hidden units. Following the activation of the hidden layer neuron, its output value serves as a new input. This new input is multiplied by the same weights as before and combined with the bias of the visible layer. This process is known as either reconstruction or backward pass. The original input and the newly generated input are then compared to assess whether they match. By comparing these inputs, the restricted Boltzmann Machine can evaluate the quality of the reconstruction and adjust its parameters accordingly during the training process. This comparison helps the model learn to recreate the original input as accurately as possible\cite{hinton2012practical}.

The RBM model is an energy-based model where the joint probability distribution
is defined by its energy function as follows:

\begin{equation}
 p(v,h) =  \frac{1}{Z} exp(-E(v,h))
\label{eqn:eqn4}
\end{equation}

In this context, $Z$ represents a partition function that is calculated as the sum
of exponential terms across all possible configurations. The probability of
a visible vector can be calculated by summing the probabilities of $p(v,h)$ across all possible configurations of the hidden layer\cite{hinton2012practical}.

\begin{equation}
 p(v) =  \frac{1}{Z} \sum_h e^{-E(v,h)}
\label{eqn:eqn5}
\end{equation}

Similarly, the probability of a hidden layer configuration can be calculated by
summing the probabilities of $p(v,h)$ over all possible visible vector configurations.

\begin{equation}
 p(h) =  \frac{1}{Z} \sum_v e^{-E(v,h)}
\label{eqn:eqn6}
\end{equation}

The visible and hidden units are independent. Thus, if there are $m$ visible units and $n$ hidden units, the conditional chance of a visible unit configuration $v$ given a hidden unit configuration $h$ is:

\begin{equation}
 p(v|h) = \prod_{i=1}^m  p(v_{i},h)
\label{eqn:eqn7}
\end{equation}

And, the conditional probability of $h$ given $v$ is:

\begin{equation}
 p(h|v) = \prod_{j=1}^n  p(h_{j},v)
\label{eqn:eqn8}
\end{equation}

Using a probabilistic interpretation of the neuron activation function, we have:

\begin{equation}
 p(v_{i}=1|h) =  \sigma(a_i + \sum_{j=1}^n w_{ij}h_j)
\label{eqn:eqn9}
\end{equation}

\begin{equation}
 p(h_{j}=1|v) =  \sigma(b_j + \sum_{i=1}^m w_{ij}v_i)
\label{eqn:eqn10}
\end{equation}

where $\sigma$ denotes the logistic sigmoid as an activation function having the following general form:

\begin{equation}
 \sigma(x) = \frac{1}{1+e^{-x}}
\label{eqn:eqn11}
\end{equation}

The sigmoid function is used for binary classification, where the output is a probability between 0 and 1, indicating how likely the input belongs to a particular class. On the other hand, the softmax function is used when there are more than two
classes. It outputs a probability distribution over multiple classes.

Training restricted Boltzmann machines aims to maximize the product of the probabilities assigned to a given training set $V$ (a matrix, where each row of it is treated as a visible vector $v$).

\begin{equation}
  Max \prod_{v\in V} p(v)
\label{eqn:eqn12}
\end{equation}

Or, equivalently, maximize the expected log probability of v. The expected log
probability is the average of the logarithm of the probabilities assigned to each training example.

\begin{equation}
  Max \ \textbf{E}  \left[\sum_{v\in V} \log p(v) \right]
\label{eqn:eqn13}
\end{equation}

The contrastive divergence (CD) algorithm is most frequently used to train RBMs by optimizing the weight matrix $W$. The weight update is computed by a gradient descent process incorporating the Gibbs sampling algorithm. Gibbs sampling estimates the posterior distribution of a set of parameters given some observed data. It is utilized to generate samples from a complex probability distribution, where each variable is sequentially updated based on the current values of all other variables. This iterative process continues until the distribution reaches a steady state. Contrastive divergence then comes into play, refining the model's parameters. It does this by contrasting the updates derived from the data with those obtained from a few iterations of Gibbs Sampling. This comparison helps to align the model's distribution more closely with the actual data. This is how the contrastive divergence algorithm works:

\begin{itemize}
    \item Start with a training sample $v$ and calculate the probabilities of the hidden units. Sample a hidden activation vector $h$ from this probability distribution.
    \item Compute the outer product of $v$ and $h$, which represents the positive gradient.
    \item From $h$, generate a reconstruction $v'$ of the visible units. Then, resample the hidden activations $h'$ based on this reconstruction (Gibbs sampling step).
    \item Calculate the outer product of $v'$ and $h'$, representing the negative gradient.
    \item Update the weight matrix $W$ using the positive gradient minus the negative gradient, multiplied by a learning rate.
    \item Update the biases $a$ and $b$ in a similar manner. 
\end{itemize}

\end{document}